%% file: Main.tex
\documentclass[lettersize,journal]{IEEEtran}
\IEEEoverridecommandlockouts

\usepackage{graphicx}
\usepackage{subfigure}
\usepackage{amsmath,amsthm,amsfonts,amssymb}
\usepackage{cite}
\usepackage{bm}
\usepackage{bbm}
\usepackage{url}
\usepackage{array}
\usepackage{color, soul}
\usepackage{multirow}
\usepackage{booktabs}
\usepackage[table,xcdraw]{xcolor}
\usepackage{enumitem}

\theoremstyle{plain}

\newtheorem{rem}{Remark}

\usepackage{tikz}
\newcommand{\tikzxmark}{%
\tikz[scale=0.23] {
    \draw[line width=0.7,line cap=round] (0,0) to [bend left=6] (1,1);
    \draw[line width=0.7,line cap=round] (0.2,0.95) to [bend right=3] (0.8,0.05);
}}
\newcommand{\tikzcmark}{%
\tikz[scale=0.23] {
    \draw[line width=0.7,line cap=round] (0.25,0) to [bend left=10] (1,1);
    \draw[line width=0.8,line cap=round] (0,0.35) to [bend right=1] (0.23,0);
}}

\allowdisplaybreaks[4]

\begin{document}
\title{Point Cloud in the Air}
\author{
Yulin~Shao,
Chenghong Bian,
Li Yang,
Qianqian Yang,
Zhaoyang Zhang,
Deniz~G\"und\"uz
\thanks{Y. Shao is with the State Key Laboratory of Internet of Things for Smart City, University of Macau, S.A.R. He is also with the Department of Electrical and Electronic Engineering, Imperial College London, London SW7 2AZ, U.K. (e-mail: ylshao@um.edu.mo).}
\thanks{C. Bian and D. G\"und\"uz are with the Department of Electrical and Electronic Engineering, Imperial College London, London SW7 2AZ, U.K. (e-mails: \{c.bian22,d.gunduz\}@imperial.ac.uk).
}
\thanks{L. Yang, Q. Yang, and Z. Zhang are with the Department of Information Science and Electronic Engineering, Zhejiang University (e-mails: \{liyang2021,qianqianyang20,ning\_ming\}@zju.edu.cn).
}
% \thanks{
% This work was supported by the European Research Council project BEACON under grant number 677854, and by CHIST-ERA grant CHIST-ERA-20-SICT-004 (funded by EPSRC-EP/W035960/1).
% }
}

\maketitle
\begin{abstract}
Acquisition and processing of point clouds (PCs) is a crucial enabler for many emerging applications reliant on 3D spatial data, such as robot navigation, autonomous vehicles, and augmented reality. In most scenarios, PCs acquired by remote sensors must be transmitted to an edge server for fusion, segmentation, or inference. Wireless transmission of PCs not only puts on increased burden on the already congested wireless spectrum, but also confronts a unique set of challenges arising from the irregular and unstructured nature of PCs. In this paper, we meticulously delineate these challenges and offer a comprehensive examination of existing solutions while candidly acknowledging their inherent limitations. In response to these intricacies, we proffer four pragmatic solution frameworks, spanning advanced techniques, hybrid schemes, and distributed data aggregation approaches. In doing so, our goal is to chart a path toward efficient, reliable, and low-latency wireless PC transmission.
\end{abstract}


\section{Introduction}
\input{SecI}

\section{Main Challenges of Wireless PC Transmission}\label{sec:challenge}
\input{SecII}

\section{Existing Solutions and Their Limitations}
\input{SecIII}

\section{Potential Solutions and Research Directions}
\input{SecIV}

\section{Conclusion}
This paper undertook a thorough investigation into the challenges surrounding wireless PC transmission and introduced innovative solution frameworks that hold the promise of reshaping the landscape of efficient 3D spatial data transmission. Our findings illuminate several critical research directions, encompassing semantic transmission of large-scale PCs, representational compression inspired by NeRF, uplink PC aggregation, and the optimization for delay-critical applications. In view of these challenges and opportunities, our work underscores the imperative for sustained innovation and interdisciplinary collaboration to seamlessly integrate 3D spatial data into the interconnected fabric of the future.

\bibliographystyle{IEEEtran}
\bibliography{References}

% \begin{IEEEbiography}{Author 1}
% \end{IEEEbiography}

% \begin{IEEEbiography}{Author 2}
% \end{IEEEbiography}

% \begin{IEEEbiography}{Author 3}
% \end{IEEEbiography}

\end{document}

%% file: SecI.tex
A point cloud (PC) is a set of points that collectively depict a physical object or scene \cite{PCL,Draco,MPEG,HoloCast,PCV}. It encapsulates both the geometry, conveyed through the three-dimensional (3D) coordinates of the points, and various attributes associated with each point, including color, reflectance, transparency, curvature, among others. PCs offer a distinct advantage over other 3D representations by providing a highly detailed and accurate depiction of an object's shape, structure, and spatial characteristics, particularly in representing complex non-manifold geometries.

As a bridge between the physical and virtual worlds, PCs find extensive applications across various fields, such as robotics, autonomous driving, digital twin, virtual/augmented reality (VR/AR), cultural heritage, telepresence, etc. \cite{MPEG,HoloCast}. For instance, the facial recognition technology commonly found on smartphones relies on PC recognition. By utilizing 3D PC data of facial features, smartphones can accurately identify and authenticate users. In autonomous driving, PCs play a pivotal role in enhancing the perception capabilities of autonomous vehicles. By providing detailed spatial information about the surrounding environment, PCs enable accurate detection and interpretation of objects, facilitating safer navigation and decision-making. PCs have also been leveraged in cutting-edge products like the Apple Vision Pro, which incorporates PC technology to create realistic and interactive virtual environments and deliver immersive VR/AR experiences to users. 

PCs are typically captured using various sensing technologies, such as light detection and ranging (LIDAR), 3D scanning, depth sensors, and photogrammetry. LIDAR, for example, emits laser beams and measures the time it takes for the light to reflect back from objects in the environment. By scanning the surroundings and recording the 3D coordinates of objects, a LIDAR can construct PCs with high accuracy and intricate details. In dynamic environments, it becomes necessary to continuously capture multiple frames of PCs. This continuous acquisition of successive frames adds an additional temporal dimension to the 3D data, resulting in the formation of 4D PCs, also known as dynamic PCs.

\begin{figure}[t]
  \centering
  \includegraphics[width=1\columnwidth]{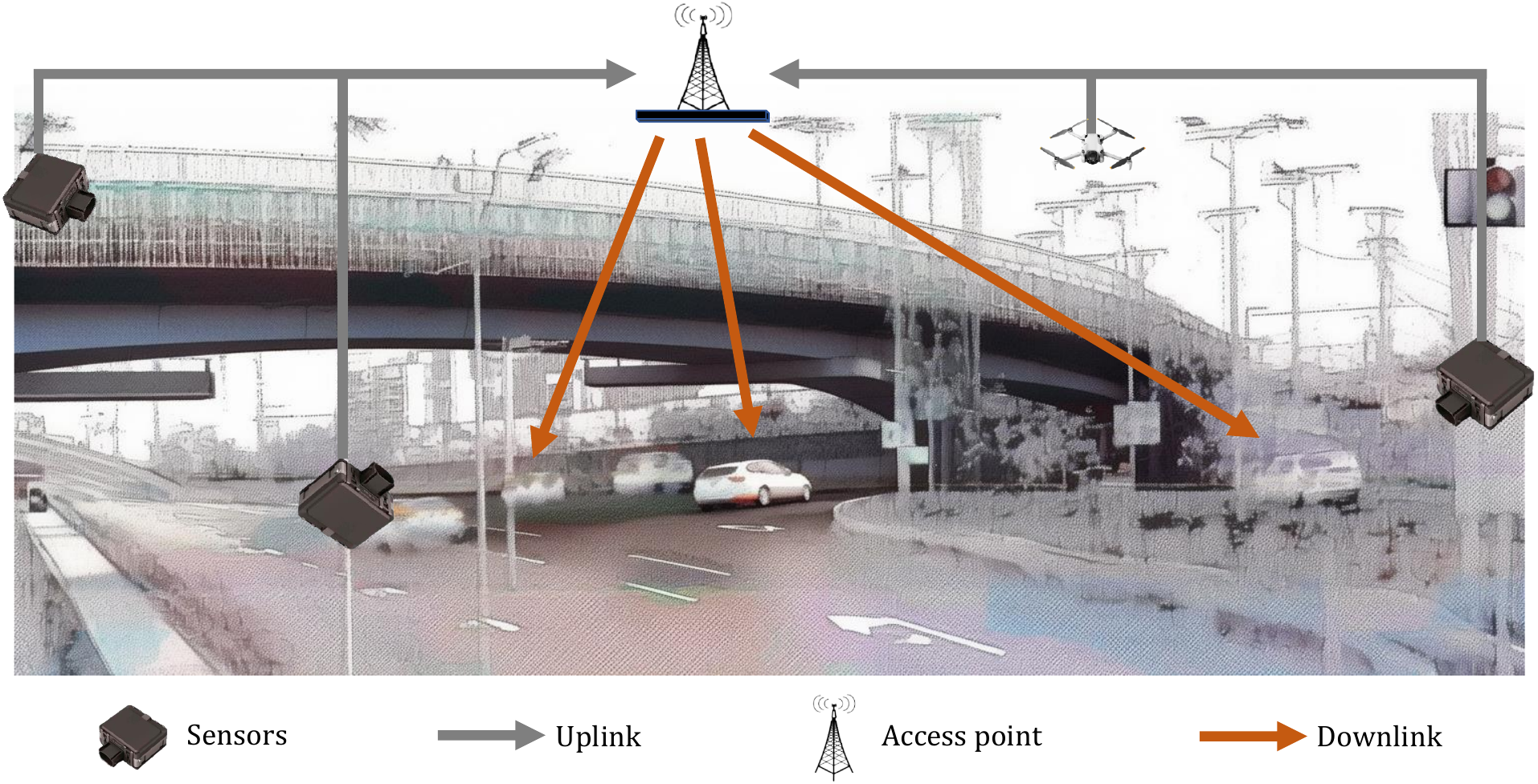}\\
  \caption{PC-aided collaborative environmental awareness for autonomous vehicles. Acquiring a comprehensive PC for a scene from distributed sensors through uplink transmission, which is then fed to users through downlink transmission.}
\label{fig:1}
\end{figure}

PCs exhibit three key characteristics \cite{PCL,MPEG,HoloCast,PCV}:

\textbf{Large data volume}. To represent complex spatial objects and scenes, PCs can comprise several million points, leading to a substantial amount of data. Considering only the representation of 3D coordinates, a PC with 1 million points requires 48Mb data. If we further consider dynamic PCs, the data volume sampled at a rate of 30 frames per second escalates to 1.44Gbps.

\textbf{Non-uniformly distributed and non-ordered}. In contrast to traditional 2D data like images or videos, PC data exhibit irregular spatial distribution, primarily caused by the shape characteristics of the objects being represented. Furthermore, the points in a PC lack inherent order, resulting in limited semantic information among these points.

\textbf{Multi-view acquisition}. The raw PCs captured from sensors exhibit another form of irregularity referred to as the ``intense in proximity, sparse in distance''. In this phenomenon, there is a higher concentration of points in close proximity, while points become sparser as the distance increases due to factors like reduced reflectivity or occlusion. Consequently, the deployment of distributed sensors for multi-view scanning, as shown in Fig. \ref{fig:1}, has become indispensable to address this irregularity and obtain a more comprehensive and detailed depiction of the environment.

To date, the study of PCs has primarily been conducted within the computer science community, with a predominant emphasis on the semantic segmentation and classification of readily available PCs. However, in practical applications, the acquisition and processing of PCs may not be co-located. Particularly in the case of multi-view acquisitions, the fusion and stitching of PCs sensed by distributed sensors are carried out at a remote server. This highlights the criticality and inevitability of wireless transmission of PCs in real-world scenarios, serving as a fundamental requirement for transitioning algorithms from laboratory experiments to practical systems. For carefully-fused high-fidelity PCs, wireless communication provides enhanced mobility and flexibility, making it essential for applications demanding real-time collaborative data processing and decision making.

In this paper, our primary focus centers on the pivotal issue of wireless transmission of PCs. Our main contributions are distilled as follows:

\begin{itemize}
    \item We emphasize the profound importance of wireless PC transmission and elucidate four major challenges it confronts. This underscores the compelling necessity for the development of dedicated wireless transmission systems tailored specifically to PCs.
    \item We conduct a comprehensive review and in-depth analysis of existing works and solutions, pinpointing their strengths and limitations. The analysis yields invaluable insights, paving the way for potential systemic solutions.
    \item We put forth four pragmatic solution frameworks in response to the intricate challenges inherent in PC transmission. These encompass a semantic communication framework grounded in deep joint source-channel coding (DeepJSCC) \cite{DeepJSCC}, a representational compression framework drawing inspiration from neural radiance fields (NeRF) \cite{Nerf}, an uplink PC feature aggregation framework, and a distributed broadcast framework tailored for applications with stringent delay constraints.
\end{itemize}

% Collectively, these contributions form a sturdy foundation for advancing the over-the-air transmission of point cloud. They not only shed light on the challenges at hand but also offer innovative pathways toward their resolution. By presenting these solution frameworks, our aim is to ignite further research and development efforts within this domain, ultimately fostering the emergence of more efficient and dependable point cloud communication systems.

%% file: SecII.tex
To start with, this section summarizes the challenges encountered by wireless PC transmission.

\textbf{Limited bandwidth.} The scarcity of spectrum in wireless communications presents a significant challenge in efficiently utilizing the available bandwidth for reliable transmission of PCs. This challenge becomes particularly prominent in scenarios involving distributed fusion, where multiple sensors need to transmit captured PCs to a common access point (AP), and dynamic PC transmission, which necessitates real-time continuous updates of spatial information.

\textbf{Efficient PC compression (PCC).} By removing redundancies in both coordinates and attributes, PCC transforms the PC to a compressed bitstream, thereby alleviating the burden of wireless transmission and reducing sensors' energy expenditure. Yet, PCs present a unique challenge because they consist of unstructured, unordered, and irregularly distributed points that often lack semantic information. The intricacy is compounded when it comes to attribute compression, as these attributes are tied to the irregular geometry. The task at hand involves not only reducing data volume but also preserving the critical information that makes PCs meaningful. Thus, the art of extracting these meaningful features and efficiently compressing PCs remains an active and challenging frontier in current research and development.

\textbf{Cliff and leveling effects.} While compression reduces the data volume, it also renders the compressed data more vulnerable to bit errors and packet losses. Mitigating these issues necessitates the adoption of channel coding.  However, two notable hurdles naturally arise: the cliff and levelling effects. The cliff effect materializes as a precipitous drop in transmission rate when the channel quality dips below a specific threshold. Meanwhile, the levelling effect embodies a scenario where the transmission rate stubbornly resists improvement despite enhancements in channel quality, unless the modulation and coding scheme (MCS) can be adaptively reconfigured in harmony with real-time channel conditions. These inherent intricacies underscore the challenge of ensuring robust and reliable PC transmission, particularly over time-varying channels.

\textbf{Delay sensitive applications.} Many PC applications, including autonomous vehicles, real-time AR/VR, remote meetings, and remote surgery, heavily depend on the seamless flow of real-time processing, visualization, and critical decision-making. In these contexts, the tolerance for delays is exceedingly low, making it imperative to strike a delicate equilibrium between the complexity and performance of compression and decompression algorithms.

%% file: SecIII.tex
Within the existing literature, considerable research efforts have been dedicated to tackling the challenges highlighted in Section \ref{sec:challenge} and facilitating effective wireless PC transmission. Table~\ref{tab:1} provides an overview of these solutions and highlights their key features. Depending on the modulation schemes employed for data transmission, these approaches can be broadly categorized as digital, analog, and hybrid schemes. In the following, we will comprehensively explore these three categories of strategies, delving into their respective merits and limitations.

\begin{table*}[]
\renewcommand*{\arraystretch}{0.9}
\setlength{\tabcolsep}{2mm} % length of the table.
\caption{A summary of existing solutions for wireless PC transmission.}
\label{tab:1}
\centering
\begin{tabular}{@{}ccccccc@{}}
\toprule
Schemes & PC data & Backbone & Modulation & lossless  & \begin{tabular}[c]{@{}c@{}}Exploit temporal\\ Correlations \end{tabular} & \begin{tabular}[c]{@{}c@{}}Prevent cliff \& \\ levelling effects\end{tabular} \\ \midrule
\cite{PCL} PCL  & Geo. \& Attr. & Octree & Digital & \tikzcmark & \tikzxmark & \tikzxmark \\
\cite{Draco} Draco & Geo. \& Attr. & KD-tree & Digital & \tikzcmark  & \tikzxmark & \tikzxmark \\
\cite{MPEG} V-PCC & Geo. \& Attr. & Video codec & Digital & \tikzcmark  & \tikzcmark & \tikzxmark \\
\cite{MPEG} G-PCC & Geo. \& Attr.& Octree & Digital & \tikzcmark  & \tikzxmark & \tikzxmark \\

\cite{Muscle} Muscle & Geo. \& Attr.  & Octree \& deep entropy model  & Digital & \tikzxmark  & \tikzcmark & \tikzxmark \\
\cite{PCAC} Deep-PCAC & Attr.  & Point Con. \& point-inception block  & Digital  & \tikzxmark  & \tikzxmark & \tikzxmark \\ 
\cite{DeepPCC} DeepPCC & Geo. \& Attr.  & Sparse conv. \& octree \& local attention & Digital  & \tikzxmark  & \tikzxmark & \tikzxmark \\
\cite{AITransfer} AITransfer & Geo. & PointNet++ & Digital & \tikzcmark  & \tikzxmark & \tikzxmark \\
\cite{PCV} PCV delivery & Geo. & PointNet++ \& PU-GAN & Digital & \tikzcmark  & \tikzxmark & \tikzxmark \\
\cite{HoloCast} HoloCast & Geo. \& Attr. & GFT & Analog & \tikzcmark  & \tikzxmark & \tikzcmark \\
\cite{HoloCast2} HoloCast+ & Geo. \& Attr. & Octree \& GFT & Hybrid & \tikzcmark  & \tikzxmark & \tikzcmark \\
\cite{SEPT} SEPT & Geo.  & Point transformer & Analog & \tikzxmark  & \tikzxmark & \tikzcmark \\

\bottomrule
\end{tabular}
\end{table*}

\subsection{Digital schemes}
Digital transmission is the classical approach for PC transmission. This method involves compressing redundancies to convert PCs into bitstreams, followed by well-established entropy coding and channel coding. The channel-coded bitstream is then modulated onto constellations for transmission. In this approach, the most challenging issue is PCC.

\begin{figure}[t]
  \centering
  \includegraphics[width=1\columnwidth]{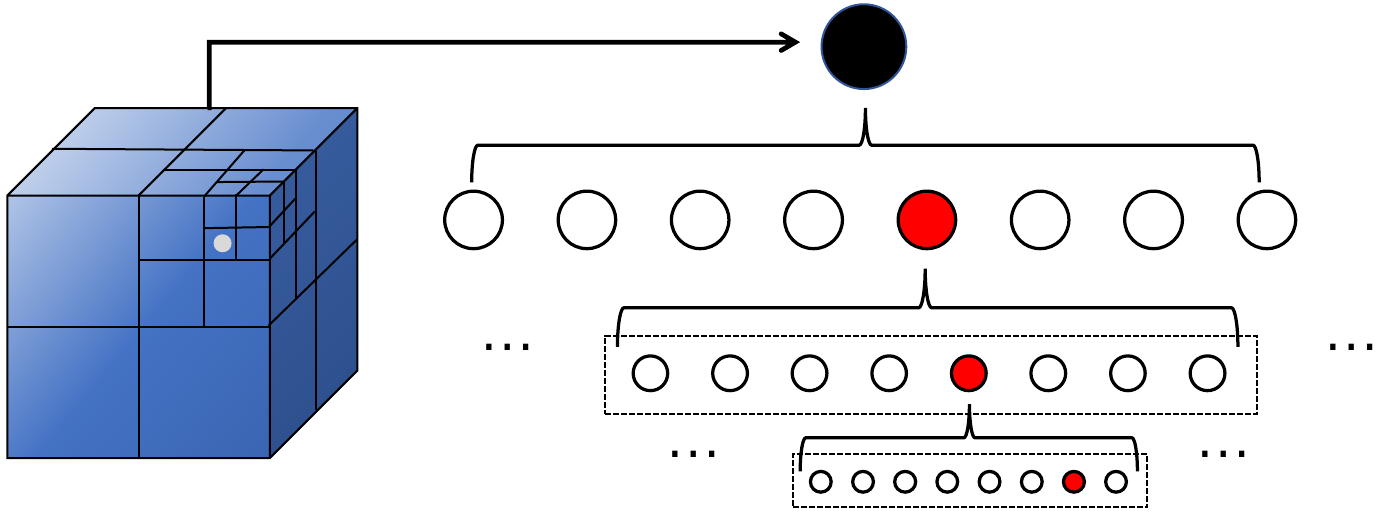}\\
  \caption{Octree encoding subdivides the 3D space into hierarchical cubic regions, storing points of the PC as leaf nodes for efficient spatial representation.}
\label{fig:2}
\end{figure}

\subsubsection{PCC standards}
As early as 2014, PCC has captured the attention of the MPEG 3D Graphics Coding Group. After thorough and iterative discussions, the first standardized PCC protocols were introduced in 2020 \cite{MPEG}. The final standards encompass two distinct compression approaches: video-based PCC (V-PCC, ISO/IEC 23090-5) and geometry-based PCC (G-PCC, ISO/IEC 23090-9).

V-PCC is primarily designed for dense PCs exhibiting a relatively uniform distribution of points. Its fundamental concept involves projecting the 3D spatial data points into a collection of 2D images, and compressing the images using 2D video codec, which is widely available on most devices nowadays. In contrast, G-PCC is designed with a focus on sparse PCs, characterized by a non-uniform distribution of points in space. As opposed to projection, G-PCC leverages the octree data structure for direct 3D space encoding. This entails dividing the space into nested cubic cells, each accommodating a specific point or set of points, as shown in Fig.~\ref{fig:2}. Both V-PCC and G-PCC support dynamic PC transmission, but only V-PCC exploits the temporal correlations among consecutive PC frames for efficient compression.

\subsubsection{Traditional approaches}
The difficulty of PCC is rooted in the irregular spatial distribution of points. To tackle this hurdle, traditional methods primarily focus on converting the irregular data structure into a more regular form. This conversion can be achieved through techniques such as projection, voxelization, or graph transformation.

Projection, as demonstrated in V-PCC, aims to reduce the 3D PCs into more manageable 2D images through multi-view projection. However, the loss of intricate 3D geometric information during projection can be a concern. Additionally, the inability to fully utilize the inherent sparsity in PCs can result in increased computational complexity and suboptimal representations.

Voxelization is a technique that discretizes 3D space into voxels, providing a structured framework for representing PCs. It often involves the use of tree structures like octrees and KD-trees. While this method has its advantages, it can introduce significant computational and memory demands, often without effectively leveraging the sparsity of PCs. Furthermore, the voxelization process itself may lead to a loss of geometric intricacies due to quantization onto a voxel grid.

Graph-based approaches prove efficient in representing complex data structures like PCs thanks to their ability to capture relationships and connections among data points. In this methodology, the geometry of PCs serves as the basis for constructing vertices and edges in a graph. Attributes associated with each data point become signals on the corresponding graph vertices. Leveraging this graph representation, we can employ graph signal processing techniques, such as graph Fourier transform (GFT) and graph wavelet transform, for efficient feature extraction and compression.

\subsubsection{Deep learning (DL)-aided approaches}
% The rapid advancement of deep learning (DL) technology has highlighted its potential in the field of PCC. 
DL enables end-to-end PCC frameworks that seamlessly encompass data transformation, feature extraction, encoding, and decoding \cite{Muscle,PCAC,DeepPCC,AITransfer,PCV}. In the following, we examine three representative works.

In \cite{Muscle}, the authors present a comprehensive approach to compress LiDAR streams by exploiting spatio-temporal redundancies through a learned deep entropy model. The approach involves quantizing and encoding spatial coordinates into an octree representation, incorporating a deep entropy model to predict occupancies and intensity values, and entropy coding for bitstream generation. They demonstrate that DL-aided schemes can efficiently compress both spatial coordinates and intensity values while preserving reconstruction quality.

Ref. \cite{PCAC} introduces Deep-PCAC, a DL-assisted approach for lossy PC attribute compression. Unlike previous methods that voxelize or project points, Deep-PCAC directly encodes and decodes attributes using geometry. It employs second-order point convolution for better spatial correlation utilization, a dense point-inception block for enhanced feature propagation, and a multiscale loss for improved optimization. While it falls short in performance compared to G-PCC, Deep-PCAC serves as a foundational work in DL-based attribute compression.

% Furthermore, DL allows us to operate directly on the irregular points without the need for any transformations \cite{DeepPCC,AITransfer,ISCom}.

Another representative PCC approach is DeepPCC \cite{DeepPCC}.
Unlike existing methods, DeepPCC offers a unified framework for both geometry and attribute compression. It leverages multiscale neighborhood information aggregation, sparse convolution, and KNN self-attention to efficiently capture spatial correlations in PCs. Moreover, DeepPCC is computationally efficient in that the computations are limited to positively-occupied voxels.

\subsection{Analog and hybrid schemes}
The theoretical foundation of digital transmission rests on Shannon's separation theorem that relies on idealized assumptions, such as ergodic sources and channels, and infinite block lengths. In practice, however, we are confronted with non-ergodic sources and channels, and operate over limited source and channel block lengths. This disparity between theoretical assumptions and real-world restrictions renders the digital approach suboptimal.
On the other hand, digital transmission suffers from the cliff and levelling effects, which become notably pronounced in time-varying and broadcast channels. Picture a typical downlink scenario in wireless networks, where an AP endeavors to broadcast a PC to multiple users. Here, the entire communication system's performance is at the mercy of its weakest link -- the channel with the poorest quality. This limitation, unfortunately, leads to missed opportunities as the potential of other, better-performing channels remain untapped.
% Similarly, even in a single receiver scenario with a time-varying channel, the transmitter has to be conservative, targeting the worst possible channel that can be encountered.

To overcome the above limitations, \cite{SEPT} proposed a new paradigm called semantic PC transmission (SEPT). Two schemes that underpin SEPT are DeepJSCC and discrete-time analog transmission, each offering distinctive advantages.
\begin{itemize}
    \item The design of JSCC systems has traditionally focused on specific sources and channels. However, DeepJSCC heralds a paradigm shift by leveraging DL to enable end-to-end learning of DeepJSCC encoder and decoder for arbitrary sources and channels as long as sufficiently rich data are available. This revolutionary approach vastly broadens the applicability of JSCC across a spectrum of scenarios.
    \item When coupled with discrete-time analog transmission, SEPT introduces a favorable feature whereby each user can reconstruct the input signal at a quality allowed by its channel noise level. This capability effectively mitigates the cliff and leveling effects. Consider the context of broadcast channels. With SEPT, the decoding performance of each user correlates directly with their individual channel quality. This eliminates the shortcoming that the overall system performance is dictated by the worst channel quality.
\end{itemize}

Another noteworthy approach that capitalizes on analog transmission is HostCast \cite{HoloCast}. In contrast to SEPT, HostCast embraces a graph-based methodology, where the PC is conceptualized as a graph, with its geometric components serving as vertices and attributes functioning as signals attributed to these vertices. HostCast leverages GFT to process these signals, subsequently transmitting them directly in an analog fashion without the intermediaries of digital quantization and entropy coding. This distinctive approach grants HostCast the capability to enhance reconstruction quality gracefully as the wireless channel quality improves.

Expanding upon HostCast, the authors further introduced HoloCast+ \cite{HoloCast2}. This iteration employs a hybrid digital-analog coding scheme, seamlessly integrating octree-based digital compression with GFT-based analog coding. This combined approach not only outperforms HostCast in terms of reconstruction quality but also effectively mitigates the cliff and leveling effects.

%% file: SecIV.tex
In the previous two sections, we have highlighted the challenges associated with wireless PC transmission and provided a concise overview of existing works. Clearly, there is a compelling need to rethink and redesign communication systems meticulously crafted for the nuances of PCs, considering their pivotal role in the impending digital landscape.
In this section, we will delve into the challenges, harness insights gleaned from the latest strides in the field, and offer our perspectives for future research.
% Our aspiration is that these proposed solutions can serve as valuable directions for future research, ultimately leading to the development of an effective wireless transmission framework uniquely suited to the characteristics of PCs.

\begin{figure}[t]
  \centering
  \includegraphics[width=0.7\columnwidth]{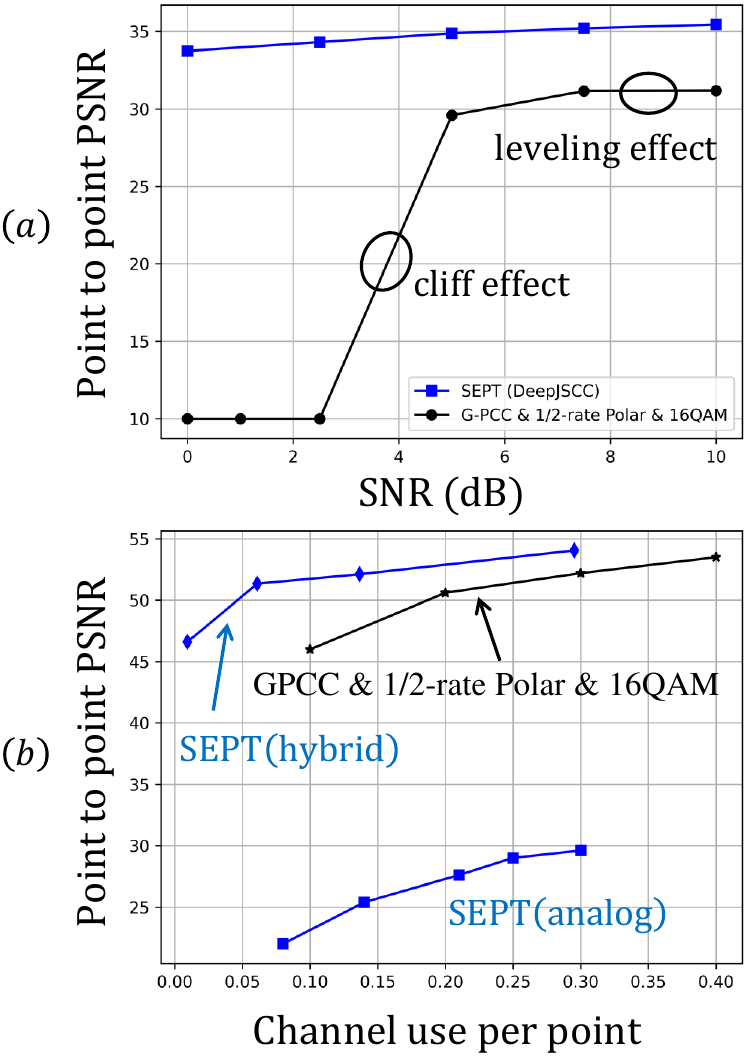}\\
  \caption{(a) SEPT efficiently addresses the cliff and leveling effects on the downsampled ShapeNet dataset. (b) The rate-distortion performance of SEPT and G-PCC on the large-scale SemanticKITTI dataset. While the raw SEPT (analog) exhibits performance degradation, the SEPT (hybrid) scheme outperforms G-PCC when transmitting certain key points to the receiver.}
\label{fig:sept}
\end{figure}

\subsection{Semantic communication}\label{sec:semantic}
Semantic communication, an emerging communication system design approach, has gained prominence in recent years \cite{semantic,DeepJSCC}. It broadly encompasses a category of end-to-end communication system designs rooted in DL.
Diverging from traditional communication systems fixated on achieving low bit error rates, semantic communication centers its focus on overall system performance and offers two main advantages.
\begin{itemize}[leftmargin=0.5cm]
    \item First, semantic communication leverages more efficient DL-based compression methods. Over recent years, DL has demonstrated superior performance in compressing various data sources, spanning voice, image, and video signals. Given the irregular nature of PCs, we foresee DL-based PCC as a major future trend and a crucial research avenue. Existing DL-based PCC endeavors \cite{PCAC,DeepPCC} have already shown advantages over traditional methods like G-PCC. 
% However, while these efforts have mostly focused on geometry compression for (dynamic) PCs, the challenge of joint geometry and attribute compression still lacks efficient solutions.
\item Second, semantic communication leverages DeepJSCC and analog transmission to mitigate the cliff and leveling effects. SEPT \cite{SEPT} represents an attempt within semantic communication for PC transmission. Leveraging point transformer's feature extraction capability and the pooling layer's feature summarization capabilities, SEPT progressively reduces the number of points within a PC, culminating in its transformation into a latent vector for analog transmission to the receiver, thereby effectively addressing the cliff and leveling effects, as shown in Fig.~\ref{fig:sept}(a).
\end{itemize}

While SEPT's success serves as a promising precedent for applying semantic communication principles to PC transmission, several challenges remain on the horizon. A key concern is that the current SEPT framework is primarily suited for small-scale PCs.
As the number of points within a PC grows, surpassing G-PCC with SEPT becomes a formidable task.
Fig.~\ref{fig:sept}(b) evaluates SEPT on a large-scale dataset. As shown, the performance of SEPT degrades significantly compared with G-PCC. This raises a pivotal question: How can we effectively employ semantic communication for large-scale PCs? Here, we propose several potential solutions:
\begin{enumerate}[leftmargin=0.5cm]
\item Craft new DL-based feature extraction and compression framework tailored for large PCs.
\item Explore hybrid digital and analog coding frameworks. In this approach, alongside transmitting the latent vector through analog means, we can also transmit key points within the large PC to the receiver in a digital format. Once these key points, which characterize the structure of the PC, are known to the receiver, the transmitter can describe the neighboring points (including both geometry and attributes) around these key points as part of the latent vector. 
Our preliminary experiments in Fig.~\ref{fig:sept}(b) indicate that this hybrid coding approach can remarkably enhance SEPT's performance on large PCs.
\item Representational compression. Another intriguing avenue is to explore representational compression methods inspired by NeRF, which we will delve into later in Section \ref{sec:NeRF}.
\end{enumerate}

Overall, the integration of DL and semantic communication holds significant promise for enhancing the efficiency and reliability of PC transmission, offering exciting opportunities for future research in this field.

\begin{rem}[Receiver with generative capabilities]
In various PC applications, the receiver often possesses a static environmental map. Leveraging this static map, the receiver can train a PC generative model, thereby gaining the ability to generate or fill in missing parts of the PCs. Exploiting this generative capability at the receiver, we can develop corresponding conditional semantic encoders at the transmitter and significantly reduce the amount of data that needs to be transmitted.
\end{rem}

\begin{figure}[t]
  \centering
  \includegraphics[width=0.7\columnwidth]{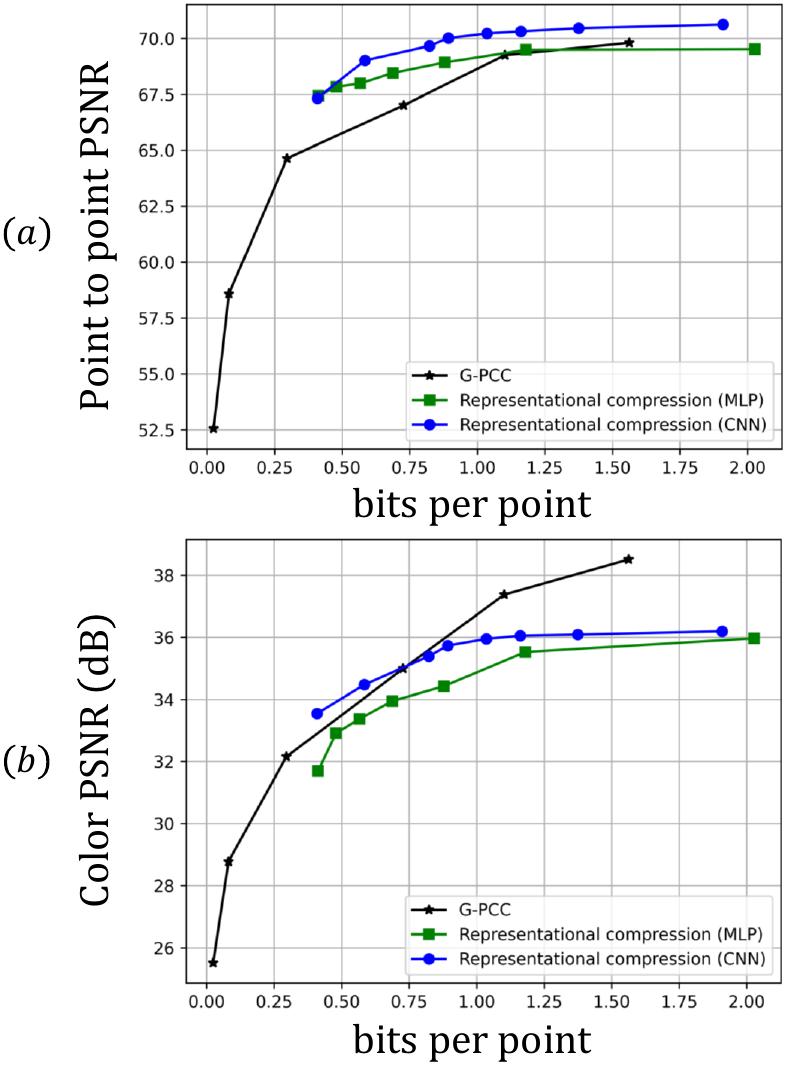}\\
  \caption{The rate-distortion performance of representational compression benchmarked against G-PCC on the 8iVFB dataset (loot). The DNN is designed to be multilayer perceptron (MLP) and CNN, respectively.}
\label{fig:nerf}
\end{figure}

\subsection{NeRF and representational compression}\label{sec:NeRF}
NeRF is an innovative approach for constructing 3D scenes and objects from 2D images \cite{Nerf}. Unlike traditional methods, NeRF takes a new route by instructing the DNN model on how light interacts with objects within a scene, effectively constructing a 3D model by capturing their appearance from multiple perspectives.

The success of NeRF not only provides a means to construct 3D PCs from 2D images but also offers an inspiring concept: when contemplating the compression of a PC, we can parametrically represent the relationship between spatial point positions and their attributes as a function (e.g., DNNs). In doing so, the transmitter only needs to transmit the parameters of this function to the receiver. The receiver can then input individual coordinates into the function to determine whether there are points at those coordinates and what attributes those points possess.
In essence, this method shifts the focus from transmitting extensive spatial data and attributes to transmitting a compact function that can efficiently represent the original PC.

When we employ DNN as the representative function, the input and output of the DNN can be written as
\begin{equation*}
(\text{Existence},\text{attributes})=f_{\bm{\theta}}(x,y,z),
\end{equation*}
where $(x,y,z)$ denotes a 3D coordinate and $f_{\bm{\theta}}$ is the DNN parameterized by $\bm{\theta}$. The training of $f_{\bm{\theta}}$ occurs at the transmitter, and subsequently, $\bm{\theta}$ is transmitted to the receiver. With $\bm{\theta}$, the receiver can deduce the presence of a point at any given position $(x,y,z)$ and retrieve associated attributes if a point indeed exists.

% As can be seen, the design of the DNN must be ingenious to ensure that transmitting $\bm{\theta}$ requires less channel use than transmitting the original PC.

The simulation results in Fig. \ref{fig:nerf} confirm the efficacy of representational compression. As shown, the DNN architecture plays a pivotal role. When we opt for a convolutional neural network (CNN), representational compression outperforms G-PCC in terms of geometry reconstruction. On the other hand, the exploration of more efficient DNN designs that can surpass G-PCC in terms of both geometry and attribute reconstruction presents an enticing research direction.

\begin{rem}[Representational compression versus semantic communication]
   When we employ DNNs as the representative functions, our challenge shifts from transmitting PCs to transmitting DNNs.
   DNN parameters are a set of real numbers. This implies that representational compression effectively transforms the PC into a latent vector. At this juncture, we can employ a semantic communication approach to transmit DNN parameters, thereby achieving higher efficiency and mitigating the cliff and leveling effects. This approach corresponds to the third solution described in Section \ref{sec:semantic} for large-scale PCs. 
\end{rem}

\subsection{Uplink PC aggregation}
So far, much of our discussion has revolved around the downlink aspect, emphasizing the efficient transmission of PCs from an AP to downlink users. Yet, another intriguing and crucial scenario is how the AP acquires PCs in the uplink. Recall that PC collection exhibits the characteristic of being intense in proximity but sparse in the distance. Therefore, an AP needs to utilize a distributed deployment of sensors to gather environmental information and construct spatial PCs. The specific solutions in practice depend on the types of sensors deployed.

When the available sensors in the environment are cameras, NeRF offers a ready-made solution. In this setup, multiple sensors transmit captured images from various perspectives to the AP. The AP then trains a DNN using these multi-view images to represent the entire 3D environment, subsequently rendering the 3D PC. From a wireless communication standpoint, the upside is that this approach can leverage well-established image encoders and decoders. The downside, however, is that it places additional responsibilities on the AP, as training the DNN and rendering the PC can consume substantial computational resources and time.

In scenarios where sensors in the environment directly collect PCs, the central challenge becomes the transmission and aggregation of distributed multi-view PCs. Unlike in point-to-point and broadcast channels, the primary objective here is to aggregate all the PCs. Thus, having each sensor transmit its PC to the AP might prove inefficient. Instead, a promising approach is feature aggregation. Here, PCs collected by multiple sensors are first transformed into feature space, with feature vectors subsequently transmitted to the AP for aggregation.
This approach raises two key questions.
First, should the transmission and aggregation of feature vectors occur separately, i.e., employing orthogonal multiple access followed by aggregation, or should they take place jointly via an over-the-air computation approach? Which approach strikes a better trade-off between feature space size and the reconstruction performance at the AP? Second, should the features extracted by multiple sensors reinforce each other or complement each other? These questions pose intriguing challenges and opportunities for optimizing the transmission of PC data in a distributed sensing environment.

%%%%%%%%%%%%%%%%%%%%%%%%%%%%%%%%%%%%%%%%%%
%%%%%%%%%%%%%%%%%%%%%%%%%%%%%%%%%%%%%%%%%%
%%%%%%%%%%%%%%%%%%%%%%%%%%%%%%%%%%%%%%%%%%
%%%%%%%%%%%%%%%%%%%%%%%%%%%%%%%%%%%%%%%%%%
%%%%%%%%%%%%%%%%%%%%%%%%%%%%%%%%%%%%%%%%%%

\begin{rem}[Multi-resolution feature aggregation]
Considering distinct channel conditions and the diverse information resolution needs for downstream tasks across various spatial directions, we can adopt a flexible multi-resolution multi-user feature aggregation approach. In this context, sensors have the liberty to downsample their PCs, with each sensor emphasizing the transmission of features at different scales. This prioritization can be determined by the significance of the transmitted information and the dynamic channel conditions. Leveraging multi-resolution feature aggregation can result in more efficient spectrum utilization without sacrificing the performance of downstream tasks.
\end{rem}

\begin{rem}[Multimodal aggregation]
Another captivating scenario arises with multimodal aggregation. In environments featuring a range of sensors like cameras for visual data and LiDAR for depth sensing, the AP has the opportunity to harness information from multiple modalities. This approach opens up exciting possibilities for integrating data from diverse sensor types, potentially leading to richer and more accurate representations of the environment. It also brings forth the challenge of efficiently transmitting and fusing data from these different sources, which is a compelling area for future exploration.
\end{rem}

\begin{figure}[t]
  \centering
  \includegraphics[width=0.68\columnwidth]{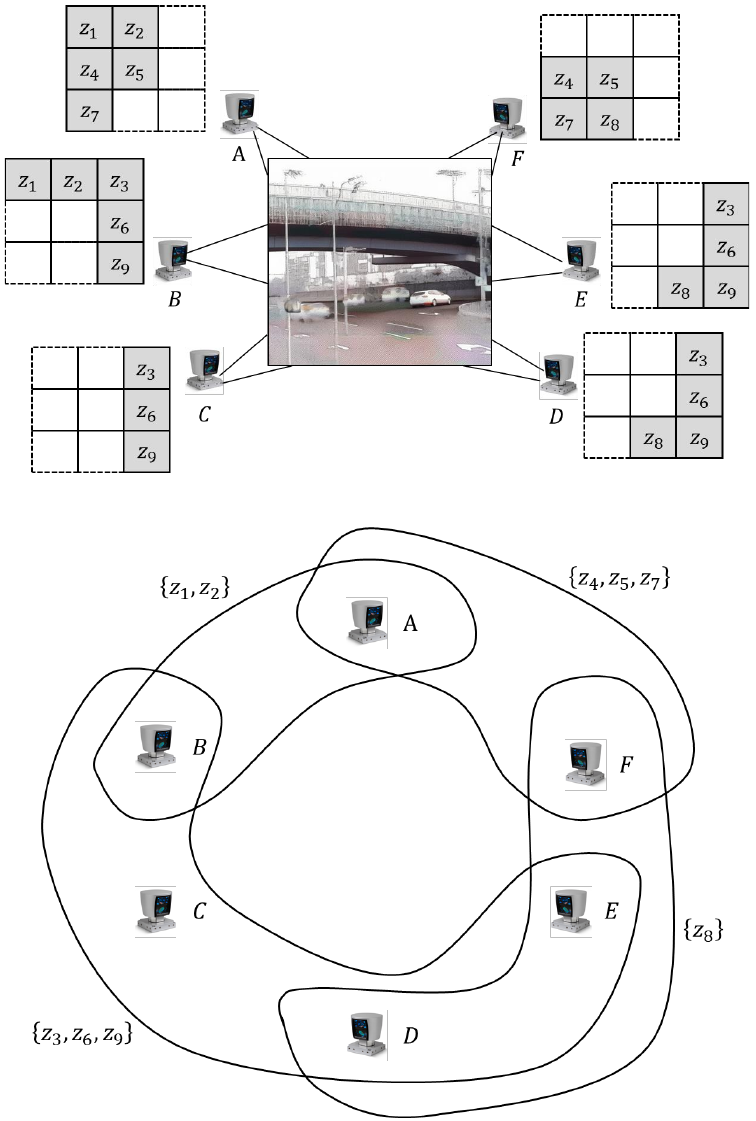}\\
  \caption{A hypergraph approach for the distributed broadcasting of PCs.}
\label{fig:hypergraph}
\end{figure}

\subsection{Delay critical applications}
In the previous analysis and discussion, we have not yet incorporated the critical consideration of delay constraints. In practical applications, information freshness stands as a paramount metric.  Outdated data loses its relevance, rendering delay a critical concern in wireless PC transmission.

PC transmission involves three main steps: compression, air interface transmission, and reconstruction. Among these, air interface transmission typically consumes the least amount of time. More often than not, compression and reconstruction are the most time-intensive steps \cite{PCV}. Thus, for applications with stringent time constraints, particularly those necessitating real-time responses, devising lightweight compression or JSCC schemes becomes pivotal in real-world scenarios.

In exceptionally critical, life-dependent situations, it might become necessary to forego compression and reconstruction, opting instead to directly downsample the raw PC data for transmission. That said, this choice does not imply a lack of optimization opportunities. It is important to recognize that PCs are a genuine reflection of 3D space, and a substantial portion of the PC data collected by distributed sensors often exhibits overlaps. This insight inspires us to craft a transmission strategy with the aim of minimizing the overall number of transmissions (and consequently, transmission time).

Consider a scenario in vehicular networks with $N$ vehicles equipped with LIDAR sensors. We can partition the complete PC of the space into $M$ distinct small regions or cubes. With LIDAR, each of the $N$ vehicles can sense a subset of these $M$ cubes. Our objective is to ensure that each vehicle can recover all $M$ cubes through wireless communication.

When a Roadside Unit (RSU) is present, the solution to this problem becomes relatively clear. Specifically, the communication process can be divided into two phases: the uplink and downlink. During the uplink phase, the RSU collects the data of all $M$ cubes. At this stage, the vehicles only need to transmit data for the cubes that the RSU fails to perceive.
In the downlink phase, the RSU broadcasts the entire PC to vehicles. At this juncture, the transmission problem can be modeled as an index coding problem \cite{bar2011index}, where the RSU holds $M$ files to distribute, and each vehicle possesses its own side information.

A more intricate scenario arises when the RSU is absent, and vehicles have to broadcast to each other, exchanging information in a self-organized manner to ensure that each vehicle ultimately obtains the data it could not perceive on its own. We refer to this problem as the distributed broadcasting problem.

Finding the optimal transmission scheme for the distributed broadcasting problem remains a formidable challenge. In this paper, we put forth a new hypergraph approach to address this challenge. To illustrate the concept, consider the scenario depicted in Fig. \ref{fig:hypergraph}, where $M=9$ and $N=6$. Each vehicle captures some regions of the scene, leading to overlapping observations. It is the intricate relationships between these observed regions that render this problem challenging.
Our hypergraph framework, however, offers a means to streamline this intricacy.
Within the hypergraph, vehicles are represented by vertices, while the observed regions are represented by edges. Given the potential for multiple vehicles to observe the same region, these edges can connect multiple vertices, giving rise to a hypergraph representation. Our approach not only streamlines the problem but also uncovers fresh analytical avenues. As an example, for a hypergraph $G$, a lower bound of the optimal number of transmissions $K$ is found to be
\begin{equation*}
    K \geq M - \text{Min-cut}(G).
\end{equation*}

%% file: Main.bbl
\begin{thebibliography}{10}
\providecommand{\url}[1]{#1}
\csname url@samestyle\endcsname
\providecommand{\newblock}{\relax}
\providecommand{\bibinfo}[2]{#2}
\providecommand{\BIBentrySTDinterwordspacing}{\spaceskip=0pt\relax}
\providecommand{\BIBentryALTinterwordstretchfactor}{4}
\providecommand{\BIBentryALTinterwordspacing}{\spaceskip=\fontdimen2\font plus
\BIBentryALTinterwordstretchfactor\fontdimen3\font minus
  \fontdimen4\font\relax}
\providecommand{\BIBforeignlanguage}[2]{{%
\expandafter\ifx\csname l@#1\endcsname\relax
\typeout{** WARNING: IEEEtran.bst: No hyphenation pattern has been}%
\typeout{** loaded for the language `#1'. Using the pattern for}%
\typeout{** the default language instead.}%
\else
\language=\csname l@#1\endcsname
\fi
#2}}
\providecommand{\BIBdecl}{\relax}
\BIBdecl

\bibitem{PCL}
R.~B. Rusu and S.~Cousins, ``3\textsc{D} is here: Point cloud library
  (\textsc{PCL}),'' in \emph{IEEE Int. Conf. Robotics and Automation}, 2011.

\bibitem{Draco}
Google, ``Draco: 3\textsc{D} data compression,'' in \emph{Github:
  \url{https://github.com/google/draco}}, 2019.

\bibitem{MPEG}
D.~Graziosi, O.~Nakagami, S.~Kuma, A.~Zaghetto, T.~Suzuki, and A.~Tabatabai,
  ``An overview of ongoing point cloud compression standardization activities:
  video-based (\textsc{V-PCC}) and geometry-based (\textsc{G-PCC}),''
  \emph{APSIPA Trans. Signal and Inf. Proc.}, vol.~9, pp. 1--17, 2020.

\bibitem{HoloCast}
T.~Fujihashi, T.~Koike-Akino, T.~Watanabe, and P.~V. Orlik,
  ``Holo\textsc{C}ast: Graph signal processing for graceful point cloud
  delivery,'' in \emph{IEEE ICC}, 2019.

\bibitem{PCV}
Y.~Huang, Y.~Zhu, X.~Qiao, X.~Su, S.~Dustdar, and P.~Zhang, ``Towards
  holographic video communications: A promising \textsc{AI}-driven solution,''
  \emph{IEEE Comm. Mag.}, 2022.

\bibitem{DeepJSCC}
E.~Bourtsoulatze, D.~B. Kurka, and D.~G{\"u}nd{\"u}z, ``Deep joint
  source-channel coding for wireless image transmission,'' \emph{IEEE Trans.
  Cognitive Comm. Netw.}, vol.~5, no.~3, pp. 567--579, 2019.

\bibitem{Nerf}
B.~Mildenhall, P.~P. Srinivasan, M.~Tancik, J.~T. Barron, R.~Ramamoorthi, and
  R.~Ng, ``Ne\textsc{RF}: Representing scenes as neural radiance fields for
  view synthesis,'' \emph{Communications of the ACM}, vol.~65, no.~1, pp.
  99--106, 2021.

\bibitem{Muscle}
S.~Biswas, J.~Liu, K.~Wong, S.~Wang, and R.~Urtasun, ``Mu\textsc{scle}: Multi
  sweep compression of \textsc{LIDAR} using deep entropy models,''
  \emph{Advances in Neural Information Processing Systems}, vol.~33, pp.
  22\,170--22\,181, 2020.

\bibitem{PCAC}
X.~Sheng, L.~Li, D.~Liu, Z.~Xiong, Z.~Li, and F.~Wu, ``Deep-\textsc{PCAC}: An
  end-to-end deep lossy compression framework for point cloud attributes,''
  \emph{IEEE Trans. Multimedia}, vol.~24, pp. 2617--2632, 2021.

\bibitem{DeepPCC}
J.~Zhang, G.~Liu, Z.~Junteng, D.~Dandan, and M.~Zhan, ``Deep\textsc{PCC}:
  Learned lossy point cloud compression,'' \emph{arXiv preprint}, 2023.

\bibitem{AITransfer}
Y.~Zhu, Y.~Huang, X.~Qiao, Z.~Tan, B.~Bai, H.~Ma, and S.~Dustdar, ``A
  semantic-aware transmission with adaptive control scheme for volumetric video
  service,'' \emph{IEEE Trans. Multimedia}, 2022.

\bibitem{HoloCast2}
T.~Fujihashi, T.~Koike-Akino, T.~Watanabe, and P.~V. Orlik, ``Holocast+: Hybrid
  digital-analog transmission for graceful point cloud delivery with graph
  fourier transform,'' \emph{IEEE Trans. Multimedia}, vol.~24, pp. 2179--2191,
  2021.

\bibitem{SEPT}
C.~Bian, Y.~Shao, and D.~Gunduz, ``Wireless point cloud transmission,''
  \emph{arXiv:2306.08730}, 2023.

\bibitem{semantic}
H.~Xie, Z.~Qin, G.~Y. Li, and B.-H. Juang, ``Deep learning enabled semantic
  communication systems,'' \emph{IEEE Trans. Signal Proce.}, vol.~69, pp.
  2663--2675, 2021.

\bibitem{bar2011index}
Z.~Bar-Yossef, Y.~Birk, T.~Jayram, and T.~Kol, ``Index coding with side
  information,'' \emph{IEEE Trans. Inf. Theory}, vol.~57, no.~3, pp.
  1479--1494, 2011.

\end{thebibliography}
